\documentclass[conference]{IEEEtran}
\IEEEoverridecommandlockouts

\usepackage{cite}
\usepackage{amsmath,amssymb,amsfonts}
\usepackage{algorithmic}
\usepackage{graphicx}
\usepackage{textcomp}
\usepackage{xcolor}
\def\BibTeX{{\rm B\kern-.05em{\sc i\kern-.025em b}\kern-.08em
    T\kern-.1667em\lower.7ex\hbox{E}\kern-.125emX}}

\usepackage{url}
\usepackage[hidelinks]{hyperref}
\usepackage{tabularx}
\usepackage{booktabs}
\usepackage{multirow}
\usepackage{subcaption}
\usepackage{pgfplots}
\usepackage{pgfplotstable}
\pgfplotsset{compat=1.18}

\usepackage{tcolorbox}
\tcbuselibrary{listings,skins}
\usepackage[font=small,skip=2pt, aboveskip=2pt, belowskip=2pt]{caption}

\begin{document}

\title{Phoneme-First Prediction for LLM-Based Speech Recognition
\thanks{Research supported by Research Foundation Flanders (FWO) under grant S004923N of the SBO programme and by the Flemish Government under the ``Flanders AI Research Program".}}

\author{\IEEEauthorblockN{1\textsuperscript{st} Jakob Poncelet}
\IEEEauthorblockA{\textit{Department Electrical Engineering ESAT-PSI} \\
\textit{KU Leuven}\\
Leuven, Belgium \\
jakob.poncelet@esat.kuleuven.be}
\and
\IEEEauthorblockN{2\textsuperscript{nd} Hugo Van hamme}
\IEEEauthorblockA{\textit{Department Electrical Engineering ESAT-PSI} \\
\textit{KU Leuven}\\
Leuven, Belgium \\
hugo.vanhamme@esat.kuleuven.be}
}

\maketitle

\begin{abstract}
Recent research has explored integrating Large Language Models (LLMs) with speech encoders to create speech-augmented LLMs capable of contextualized speech recognition. The main challenge lies in aligning the semantic embeddings of LLMs with the acoustic representations of speech encoders. We propose a novel approach that teaches the LLM to first predict phonemes from the speech features before generating the final transcript. By integrating a phoneme prediction step directly into the LLM, the model develops a fine-grained knowledge of pronunciation, reducing acoustic confusion and improving transcription accuracy and explainability. Our method is cheap and simple, as phoneme targets can be automatically derived from existing transcripts. Through comprehensive experiments, we show that intermediate phoneme prediction can improve speech recognition, particularly in low-resource settings, and yields outputs that are acoustically more faithful to the speech.
\end{abstract}


\section{Introduction}
In recent years, Automatic Speech Recognition (ASR) has made big steps forward by leveraging sophisticated sequence-to-sequence methods for attention-based models and massive data \cite{watanabe2017_jstsp, radford23_icml, peng23_asru}. The top systems rely on self-supervised learning (SSL) with large amounts of unlabeled speech data \cite{baevski20_neurips, hsu2021_taslp} followed by task-specific fine-tuning, or on supervised training with web-scraped transcripts and labeled speech datasets \cite{radford23_icml, peng23_asru}.
Similarly, Large Language Models (LLM) have seen tremendous improvements in capabilities through self-supervised pre-training of giant decoder-only architectures on textual data \cite{gpt3_neurips}. 
LLMs use context and memory to achieve strong results with prompting and in-context learning \cite{chen-etal-2025-icleval}. 

Recently, there has been a growing interest in creating ``speech-augmented LLMs'', i.e., extending LLMs with speech encoders to recognize and/or understand audio \cite{fathullah-etal-2024-audiochatllama, fathullah24_icassp, tang24_iclr, wu23_asru, deng-etal-2025-wav2prompt, seide24_interspeech, chen24_icassp}. If such speech understanding is sufficiently optimized, the LLM could transfer a wide range of its abilities to the speech domain, such as direct speech translation \cite{huang24h_interspeech}, summarization \cite{kang24_interspeech} or contextual biasing \cite{gong24b_interspeech, yang24f_interspeech}.
The integration of speech features in the LLM can be based on discrete speech units \cite{zhang-etal-2023-speechgpt}, cross-attention layers \cite{li24b_interspeech, radhakrishnan-etal-2023-whispering}, or a direct projection of continuous speech features into the LLM's hidden space \cite{wang23_icassp, wu23_asru, fathullah-etal-2024-audiochatllama}. In the latter and most popular case, the speech features are aligned and converted to the LLM's embedding space with an adapter layer \cite{wang23_icassp, yu24_icassp}. 

In general, the alignment and projection between the speech model features and the text-based LLM is learned through the task of speech recognition \cite{ma25_aaai}, using labeled speech data. The LLM is typically prompted to \textit{``Transcribe the speech to text''}, followed or preceded by the speech features. However, there is a discrepancy between the textual LLM embeddings, which are related to semantics and concepts (e.g. \textit{bank - money - economy}), and the speech encoder outputs, which are more related to pronunciation (e.g. \textit{bank - rank - dank}). 
Additionally, homophonic heterographs (e.g. \textit{cite - sight}) and homographs (e.g. \textit{cough - though}) present challenges for speech-text alignment.
Using ASR data, the LLM can learn the relation between the acoustic features and the pronunciation of words, but this is limited by the number of examples. 

We propose to teach the LLM to first predict the phonemes that it recognizes in the speech features, and then the words. By explicitly modeling both phonemes and words, the LLM attains a deeper and more fine-grained knowledge of the pronunciation of words and interpretation of the speech features. Moreover, it facilitates learning the relation between similar sounding words, even with limited amounts of speech data. Phoneme classification was also found useful in spoken language modeling to improve lexical comprehension with small amounts of data \cite{poli-etal-2024-improving}, and to improve acoustic biasing \cite{qiu23_asru, futami24_icassp}. Finally, if the LLM is prompted to predict both the phonemes spoken and the full transcript, the phonemes offer some explainability to the user in which sounds were actually discovered and why a specific word is predicted.

In this paper, we show that modeling and predicting phonemes first improves LLM-based speech recognition over the standard approach. These phonemes can be automatically generated from the speech transcripts, hence no additional manual labeling is required. Moreover, we show that our method results in transcripts that are acoustically closer to the true speech. We carry out several experiments on a variety of datasets and models, showcasing the broad applicability of our method. A relatively small amount of labeled data suffices for the LLM to learn speech patterns and pronunciations.

\begin{figure*}[h]
\centering
\begin{tcolorbox}[
    colback=gray!5!white, 
    colframe=gray!60!black, 
    sharp corners, 
    boxrule=0.5pt, 
    leftrule=1pt, 
    rightrule=1pt,
    width=\textwidth,
    title=\begin{tabular}{@{}m{0.38\textwidth} | m{0.62\textwidth}@{}}
            \centering \textbf{S2T} & \centering \textbf{Phoneme-First S2T (PF-S2T)} 
           \end{tabular},
    fonttitle=\bfseries,
    top=1mm,
    bottom=1mm,
    boxsep=1mm,
]

\small
\begin{tabular}{@{}m{0.38\textwidth} | m{0.62\textwidth}@{}}
\verb|<speech>| Transcribe the previous speech to text. &
\verb|<speech>| Transcribe the previous speech first to phonemes and then to text. \\
\quad $\rightarrow$ \textit{Transcription: Hello guys.} &
\quad $\rightarrow$ \textit{Phonemic transcription: HH AH0 L OW1 G AY1 Z.} \\ 
& \quad $\rightarrow$ \textit{Transcription: Hello guys.} \\
\end{tabular}
\end{tcolorbox}
\caption{Example of input prompt and \textit{target response} for standard speech-to-text (S2T) training and the proposed phoneme-first method.}
\label{fig:model}
\end{figure*}

\section{Speech-augmented LLM} \label{sec:sallm}
\noindent Speech-augmented LLMs consist of three parts: 1) a speech encoder, 2) a projection layer, and 3) a backbone LLM. 
Each of these blocks can be chosen to be further optimized for speech tasks \cite{ma25_aaai}.

First, the speech input is transformed to a sequence of feature vectors by a speech encoder. The speech encoder is typically pre-trained on ASR tasks \cite{wang23_icassp} or self-supervised objectives \cite{hu-etal-2024-wavllm}. We use speech encoders from strong open-source foundation models that are pre-trained on large amounts of speech data, such as the encoder from Whisper \cite{radford23_icml} or HuBERT \cite{hsu2021_taslp}. The speech encoders are kept frozen at all times. 

Second, the projection layer maps the speech features to the embedding space of the LLM.
We adopt the architecture of \cite{ma25_aaai}, which consists of an MLP network preceded by stacking of the input features.

Third, the projected speech features are fed directly into the LLM. We build the LLM prompt with the speech features first, followed by a textual instruction \cite{fan25_jstsp}. 
The text instruction is tokenized and embedded by the LLM's input layer. We perform efficient adaptation of the LLM using 4-bit QLoRA \cite{dettmers23_neurips} to further improve the modeling of speech features in the LLM by allowing small low-rank adaptations of the attention and feed-forward layers \cite{fathullah24_icassp, tang24_iclr, wu23_asru}. 

Finally, the entire pipeline is trained end-to-end with the next-token (Cross-Entropy) prediction loss of the LLM, optimizing the LoRA weights and the projection layer jointly.

\subsection{Speech-to-Text (S2T)}
\noindent The standard approach to train speech-augmented LLMs for ASR uses a simple speech-to-text instruction prompt \cite{wu23_asru, yu24_icassp, ma25_aaai, chen24_icassp, wang23_icassp}, i.e. \textit{``Transcribe the previous audio to text.''} As such, it only focuses on learning the mapping from speech features to text. We refer to this prompt and training as \textit{S2T}.

\subsection{Phoneme-First Speech-to-Text (PF-S2T)} \label{sec:pfs2t}
\noindent We propose to improve the acoustic awareness of the LLM by first predicting the phonemes and then the text. We alter the prompt to \textit{``Transcribe the previous audio first to phonemes and then to text.''} and train the model to generate the following format: \textit{``Phonemic transcription: \{phonemes\}. Transcription: \{text\}.''} We refer to this prompt as ``phoneme-first'', or \textit{PF-S2T}. Figure~\ref{fig:model} shows an example. 

Analysis of the LLM's attention maps showed that while the LLM attended to the speech features to predict the phonemes, it almost only looked at the phonemes to generate the transcription. This is a consequence of teacher-forcing training in decoder-only LLMs, i.e. the LLM is always given the ground-truth tokens as input. Therefore, we propose a joint training scheme, which randomly chooses between the speech-to-text instruction and the phoneme-first instruction prompts. We refer to this prompting technique as \textit{Joint}. During inference, we can either set the \textit{S2T} prompt or the \textit{PF-S2T} prompt.

\section{Setup}
\noindent Unless otherwise specified (e.g. a different encoder), the following setup is used throughout the paper.

\textbf{Speech encoder}: We use the \textit{Whisper-medium} encoder (24 layers, 1024 dim) \cite{radford23_icml} and stack 5 consecutive features \cite{ma25_aaai} for a feature rate of 10~Hz, padding the last vector if needed.

\textbf{Projection layer:} The stacked features are projected to a hidden dimension of 2048 (Linear + ReLU) and then to the LLM dimension of 4096, totaling 18.6M parameters.

\textbf{LLM:} We fine-tune \textit{Llama-3.1-8B} using 4-bit quantization and QLoRA \cite{dettmers23_neurips} on all linear layers (attention and feed-forward) with rank 4, alpha 16, and 0.1 dropout, resulting in 10.7M trainable parameters.

\textbf{Optimization:} We use a quantized 8-bit Adam optimizer, gradient checkpointing and Flash Attention optimization. The choice for medium-sized speech models and 4-bit QLoRA is driven by compute constraints, and they allow training the whole pipeline on a 24GB GPU card (RTX 3090).
The effective batch size is 128 or 256 utterances depending on the dataset (30 minutes of audio). We use a linear decaying learning rate schedule with 10\% warm-up ratio, a max learning rate of 5e-5 and a weight decay factor of 0.01. 

\textbf{Inference:} We use LLM beam search with a beam size of 10. The initial search is deterministic (no sampling), i.e. always selecting the tokens with the highest probability. We determine if a result is valid by checking if a specific prompt word is in the hypothesis (e.g. ``\textit{Transcription:}'' in PF-S2T) and if the compression ratio\footnote{Defined as the original UTF-8 byte length divided by the compressed byte length using \textit{zlib}.} of the prediction is below 2.4 to detect hallucinations \cite{radford23_icml}. If the best result is not valid, we look for alternatives in the rest of the beam. If there are no valid predictions in the whole beam, we do generation again with sampling enabled and increase the final softmax temperature by 0.2 until a valid response is found.

\textbf{Scoring:} We report error rates after normalization of the predicted and reference transcripts, such as lowercasing, removing punctuation and digits (or converting numbers to text for Whisper).

\begin{table}[t]
\centering
\footnotesize
\caption{\textbf{LibriSpeech-100h \& TEDLIUM-100h}: WER (\%) when training on 100h of LibriSpeech (\textit{train-clean-100}) or TED-LIUM.}
\label{tab:lib100}
\resizebox{\columnwidth}{!}{%
\begin{tabular}{l l l | c c c c c | c}
\toprule
\multicolumn{3}{c}{\textbf{Setup}} & \multicolumn{5}{c}{\textbf{WER  - LibriSpeech}} & \textbf{TED} \\
\multirow{2}{*}{\textbf{Model}} & \multirow{2}{*}{\textbf{Train}} & \multirow{2}{*}{\textbf{Decode}} & \textbf{\textit{dev-}} & \textbf{\textit{dev-}} & \textbf{\textit{test-}} & \textbf{\textit{test-}} & \multirow{2}{*}{\textbf{\textit{AVG}}} & \multirow{2}{*}{\textbf{\textit{test}}}\\
& & & \textbf{\textit{clean}} & \textbf{\textit{other}} & \textbf{\textit{clean}} & \textbf{\textit{other}} & & \\
\midrule
Whisper-M & -- & Beam & 6.5 & 11.3 & 7.3 & 12.2 & 9.3 & 15.6 \\
\midrule
\multirow{4}{*}{\shortstack{Whisper-M\\+ Llama 8B\\(4bit, r=4)}} & S2T & S2T & 5.9 & \textbf{8.4} & 5.6 & 9.4 & 7.3 & 8.9 \\
\cline{2-9}
& PF-S2T & PF-S2T & 5.0 & 9.9 & 5.0 & 10.5 & 7.6 & \underline{7.0} \\
& Joint & S2T & \underline{4.8} & 9.1 & \underline{4.8} & \textbf{9.2} & \underline{7.0} & 9.3 \\
& Joint & PF-S2T & \textbf{4.2} & \underline{8.9} & \textbf{4.6} & \underline{9.3} & \textbf{6.8} & \textbf{6.7} \\
\bottomrule
\end{tabular}
}
\end{table}

\section{Experiments}
\subsection{Evaluation on 100h of LibriSpeech and TED-LIUM} \label{sec:100h}
\noindent We investigate the effectiveness of our phoneme-first approach and compare to  the standard S2T method. We split our experiments based on the size and type of the training data.

\subsubsection{\textbf{LibriSpeech-100h}} \label{sec:ls100h}
\noindent First, we train on the \textit{train-clean-100} split of LibriSpeech. We hold out 1024 random utterances for validation. Phonetic transcripts for LibriSpeech have been generated and aligned with Montreal Forced Aligner and are publicly available \cite{lugosch19_interspeech}. The aligner uses the ARPAbet phoneme set and was trained on the full LibriSpeech dataset. We report the Word Error Rates (WER) in Table~\ref{tab:lib100}. For reference, we add the result of zero-shot Whisper-medium which uses the same speech encoder (with careful text normalization).

\subsubsection{\textbf{TEDLIUM-100h}}
\noindent We replicate the setup on 100h of TED-LIUM (v3) \cite{hernandez2018_specom}. We use the provided dictionary (an extension of CMUdict) to map words to phonemes, relying on the pre-computed forced alignment to select the correct pronunciation variant for each word \cite{rousseau-etal-2012-ted}. The phonemes follow the ARPAbet format. We test on the official test split (3h of speech). Results are also in Table~\ref{tab:lib100}. 

\subsubsection*{\textbf{Discussion of Results}}
Joint training with phoneme-first S2T improves the WER for nearly all test sets. For LibriSpeech, we remark very strong improvements for the \textit{clean} test sets, likely due to the fact that the phoneme models are trained on clean data only and thus have more difficulty recognizing phonemes for the data in the \textit{other} test sets. For TED-LIUM, we find substantial improvements of up to 25\% over S2T using phoneme-first S2T, potentially due to the quality of the phoneme labels and the more spontaneous nature.

\subsection{Evaluation on 960h of LibriSpeech} \label{sec:ls960h}
\noindent To investigate whether the phoneme-first method scales with more labeled data, we repeat the Section~\ref{sec:ls100h} experiment using the full 960h LibriSpeech set. In this data rich setting, we increase the LLM QLoRA rank to 32 with alpha 64 and dropout 0.05. We will perform two analyses with different encoders.

\subsubsection{\textbf{Whisper Encoder}} First, we use the encoder from Whisper-large-v3 (to get more competitive results compared to Whisper-medium), which has 32 layers and an output dimension of 1280. This setup results in 52M trainable parameters. Results are in Table~\ref{tab:lib960}. For reference, we add zero-shot Whisper-large results.

\subsubsection{\textbf{SSL Encoder}}
We explore whether phoneme-first decoding is effective with self-supervised (SSL) speech encoders, as they tend to be strong encoders of phonetic properties \cite{pasad21_asru}, yet they are not trained with textual data. We use the HuBERT large encoder \cite{hsu2021_taslp}, which was pre-trained on 60k hours of LibriLight (24 layers, 1080 dim). Similar to \cite{yang21c_interspeech}, we implement a learnable weighted sum of the hidden states of all layers, since some non-final SSL layers are more optimal for phoneme and/or word recognition \cite{pasad21_asru}. These layer weights are jointly optimized with the projection layer and the LLM, totaling 50M parameters. Results are also in Table~\ref{tab:lib960}. For reference, we add the result of a fully finetuned HuBERT-Large encoder using CTC loss on LibriSpeech 960h (300M trainable parameters).

\subsubsection*{\textbf{Discussion of Results}} 
Table~\ref{tab:lib960} shows significant improvements with joint training of S2T and PF-S2T, which establishes strong connections between pronunciations and speech transcripts in the LLM. Additionally, we notice that the self-supervised encoder with weighted layer sum tends to outperform the supervised Whisper encoder on this in-domain speech task. While for 100h of speech data (Table~\ref{tab:lib100}) PF-S2T decoding was optimal, in larger data regimes S2T decoding tends to be better (as it has seen enough examples during training). Still, joint training with PF-S2T brings strong benefits. We remark that for both encoder types, the fully trained baseline is still slightly better than the LLM model (as is generally the case), which could probably be equalised by using the LLM in full precision, larger rank, or more diverse speech data.

\begin{table}[t]
\centering
\footnotesize
\caption{\textbf{LibriSpeech-960h}: WER (\%) when training on full 960h of LibriSpeech.}
\label{tab:lib960}
\resizebox{\columnwidth}{!}{%
\begin{tabular}{l l l | c c c c c}
\toprule
\multicolumn{3}{c}{\textbf{Setup}} & \multicolumn{4}{c}{\textbf{WER}} & \\
\multirow{2}{*}{\textbf{Model}} & \multirow{2}{*}{\textbf{Train}} & \multirow{2}{*}{\textbf{Decode}} & \textbf{\textit{dev-}} & \textbf{\textit{dev-}} & \textbf{\textit{test-}} & \textbf{\textit{test-}} & \multirow{2}{*}{\textbf{\textit{AVG}}}\\
& & & \textbf{\textit{clean}}& \textbf{\textit{other}} & \textbf{\textit{clean}} & \textbf{\textit{other}} & \\
\midrule
Whisper-L & -- & Beam & 2.4 & 3.9 & 2.3 & 4.1 & 3.2 \\
\hline
\multirow{4}{*}{\shortstack{Whisper-L\\+ Llama 8B\\  (4bit, r=32)}} & S2T & S2T & 3.1 & 5.7 & 3.3 & \underline{5.6} & 4.4 \\
\cline{2-8}
& PF-S2T & PF-S2T & 3.1 & 5.5 & 2.8 & 6.1 & 4.4 \\
& Joint & S2T & \textbf{2.3} & \underline{5.0} & \textbf{2.2} & \textbf{5.1} & \textbf{3.7} \\
& Joint & PF-S2T & \underline{2.8} & \textbf{4.9} & \underline{2.6} & 5.9 & \underline{4.1} \\
\bottomrule
\toprule
HuBERT-L & CTC & Greedy & 1.9 & 4.0 & 2.0 & 4.1 & 3.0 \\
\hline
\multirow{4}{*}{\shortstack{HuBERT-L\\+ Llama 8B\\  (4bit, r=32)}} & S2T & S2T & 2.8 & 4.9 & 2.8 & 5.3 & 4.0 \\
\cline{2-8}
& PF-S2T & PF-S2T & 3.0 & 4.2 & 2.6 & 5.6 & 3.9 \\
& Joint & S2T & \textbf{2.0} & 4.2 & \textbf{2.0} & \textbf{4.4} & \textbf{3.2} \\
& Joint & PF-S2T & \underline{2.6} & \textbf{3.6} & \underline{2.2} & \underline{4.9} & \underline{3.3} \\
\bottomrule
\end{tabular}
}
\end{table}

\subsection{Impact of Phoneme Label Quality} \label{sec:qual}
\noindent In this section we evaluate the impact of the quality of the phoneme labels on our phoneme-first method. 
We use the Spoken Dutch Corpus \cite{CGN_Oostdijk} which contains 240h of labeled Flemish Dutch speech and automatically generated phoneme labels resulting from forced alignment with multiple pronunciations per word \cite{demuynck02_icslp}. For a subset of 25h, there are also manually corrected phoneme labels \cite{goddijn2003}. We apply our method to 1) only the data with manual phoneme labels, 2) all data using only automatic phoneme labels, and 3) a combination of both (i.e. using the manual phoneme labels when available). 

We use a Conformer speech encoder pre-trained on 14k hours of weakly supervised Belgian Dutch speech \cite{poncelet2025} in an encoder-decoder model with CTC regularization \cite{watanabe2017_jstsp}, and the Dutch \textit{Tweety-7B} LLM \cite{remy24_colm} which is adapted from \textit{Mistral-7B}. Table~\ref{tab:cgn} reports WERs against transcripts and PERs against manually labeled phonemes in the test set, evaluated on a test set with 4k utterances (3 hours). 

\begin{table}[t]
    \centering
    \footnotesize
    \caption{\textbf{Spoken Dutch Corpus}: WER (\%) and PER (\%) with Tweety LLM and Dutch fine-tuned ASR encoder, using either manually labeled phoneme data or automatically labeled phoneme data.}
    \label{tab:cgn}
    \begin{tabular}{p{2cm} l l | c c}
        \toprule
        \textbf{Phoneme Labels} & \textbf{Training} & \textbf{Decoding} & \textbf{WER} & \textbf{PER} \\
        \midrule
        \multirow{4}{*}{Manual (25h)} & S2T & S2T & 17.0 & -- \\
        & PF-S2T & PF-S2T & 12.9 & 8.3 \\
        & Joint & S2T & 14.4 & -- \\
        & Joint & PF-S2T & \textbf{11.4} & 8.5 \\
        \midrule
        \multirow{4}{*}{Automatic (240h)} & S2T & S2T & 13.1 & -- \\
        & PF-S2T & PF-S2T & 11.0 & 10.0 \\
        & Joint & S2T & 12.6 & -- \\
        & Joint & PF-S2T & \textbf{10.2} & 11.1 \\
        \midrule
        \multirow{4}{*}{Combined (240h)} & S2T & S2T & 13.1 & --\\
        & PF-S2T & PF-S2T & 10.5 & 10.0 \\
        & Joint & S2T & 10.5 & -- \\
        & Joint & PF-S2T & \textbf{10.0} & 10.5 \\
        \bottomrule
    \end{tabular}
\end{table}

The results indicate that manually labeled phonemes (i.e. higher quality labels) are more beneficial for accurate phoneme recognition, which in turn is favorable for the phoneme-first method, and that the method also scales with automatic phoneme labels. 
In all cases, phoneme-first S2T outperforms S2T training. 
PF-S2T attains strong PERs, but tends to overfit on the predicted phonemes for the final words (cf. Section~\ref{sec:pfs2t}), leading to higher WERs than Joint training.

\subsection{Analysis of Speech-Text Alignment} \label{sec:ana}
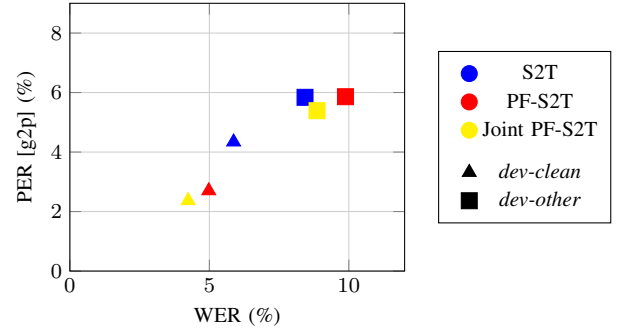
\begin{figure}[t]
    \centering
    \footnotesize
    \begin{tikzpicture}
    \begin{axis}[
        xlabel={WER (\%)},
        ylabel={PER [g2p] (\%)},
        xlabel style={font=\footnotesize},
        ylabel style={font=\footnotesize},
        scale only axis,
        grid=both,
        grid style={line width=0.3pt, draw=gray!40},
        width=0.5\linewidth, 
        height=0.4\linewidth,
        mark options={scale=1.5},
        xmin=0, xmax=12.0,
        ymin=0, ymax=9.0,
        legend columns=1, 
        legend style={
            at={(1.1, 0.5)},
            anchor=west,
            font=\footnotesize,
            inner sep=3pt,
            nodes={inner sep=2pt},
        }
    ]
    \addlegendimage{only marks, mark=*, blue}
    \addlegendentry{S2T}
    \addlegendimage{only marks, mark=*, red}
    \addlegendentry{PF-S2T}
    \addlegendimage{only marks, mark=*, yellow}
    \addlegendentry{Joint PF-S2T}
    \addlegendimage{empty legend}
    \addlegendentry{}
    \addlegendimage{only marks, mark=triangle*, black}
    \addlegendentry{\textit{dev-clean}}
    \addlegendimage{only marks, mark=square*, black}
    \addlegendentry{\textit{dev-other}}
    \addplot[only marks, mark=triangle*, blue] coordinates {(5.87, 4.34)};
    \addplot[only marks, mark=square*, blue] coordinates {(8.43, 5.84)};
    \addplot[only marks, mark=triangle*, red] coordinates {(4.98, 2.70)};
    \addplot[only marks, mark=square*, red] coordinates {(9.88, 5.86)};
    \addplot[only marks, mark=triangle*, yellow] coordinates {(4.24, 2.37)};
    \addplot[only marks, mark=square*, yellow] coordinates {(8.86, 5.39)};
    \end{axis}
    \end{tikzpicture}
    \caption{Comparison between G2P-based PER and WER for Whisper-M + Llama-8B, trained on LibriSpeech \textit{train-clean-100}.}
    \label{fig:per}
\end{figure}

\noindent We have proposed to leverage phoneme labels to improve the recognition of fine-grained acoustic speech elements and facilitate pronunciation modeling in the LLM.
As argued, related words or embeddings in the LLM might correspond to different sounds, whereas for speech encoders and decoders similar embeddings are related to similar sounding words. To investigate whether phoneme-first decoding reduces acoustic confusion, we compute the Phoneme Error Rate (PER) between the phonemized predicted transcripts and phonemized reference transcripts. To this end, we use a North American English G2P (\textit{g2p-en} package\footnote{\url{pypi.org/project/g2p-en/}}) for both, purely based on the text. The G2P estimates part-of-speech tags to disambiguate between homophones, performs a lookup in the CMUDict (ARPAbet), and leverages a neural net for out-of-vocabulary words. Figure~\ref{fig:per} depicts the WER and the G2P-based PER for the experiments from Section~\ref{sec:ls100h}.

The plot illustrates several aspects. On \textit{dev-other} the WER of PF-S2T was slightly higher but the PER is still lower than the baseline S2T training. For \textit{dev-clean} the WER was reduced by 29\% compared to S2T training, but the PER is reduced by almost 50\%. This suggests that the generated transcripts more accurately preserve the acoustic characteristics of the speech. 

Finally, Table~\ref{tab:ex} shows some examples of improved acoustic perception by the LLM from Section~\ref{sec:ls100h}, as a result of phoneme prediction. The mistakes tend to make more sense acoustically, especially for rare and out-of-vocabulary words. We also noticed reduced hallucination as a result of grounding prediction in phonemes.

\begin{table}[t]
    \centering
    \footnotesize
    \caption{\textbf{Error analysis}: examples from LibriSpeech (1,3) and TED-LIUM test sets (2) with predicted phonemes for the short ones.}
    \label{tab:ex}
    \begin{tabular}{l p{6.8cm}}
        \toprule
        Ref. (1) & \textit{\textbf{rhein} and \textbf{moselle} eighteen ninety three} \\
        S2T & \textit{\textbf{romeo} and \textbf{juliet} eighteen ninety three} \\
        PF-S2T &\textit{\textbf{R AY1 AH0 N} AE1 N D \textbf{M OW0 S EH1 L} EY1 T IY1 N N AY1 N T IY0 TH R IY1} -- \textit{\textbf{ryan} and \textbf{mozel} eighteen ninety three} \\
        \hline
        Ref. (2) & \textit{look at their \textbf{bellies} pink they're feasting} \\
        S2T & \textit{look at their \textbf{ballots} pink they're feasting} \\
        PF-S2T & \textit{L UH K AE T DH EH R \textbf{B AE L IY Z} P IH NG K DH EY R F IY S T IH NG} -- \textit{look at their \textbf{bellies} pink they're feasting} \\
        \hline
        Ref. (3) & \textit{\textbf{lacrima christi} a still wine of excellent flavor \textbf{and bouquet}} \\
        S2T & \textit{\textbf{la creme a cristy} a still wine of excellent flavor \textbf{in beaujolais}} \\
        PF-S2T & ... -- \textit{\textbf{lacrimae christi} a still wine of excellent flavor \textbf{in bockay}} \\
        \bottomrule
    \end{tabular}
\end{table}

\section{Discussion}
\noindent Phoneme-first prediction can be applied to most labeled speech datasets and languages. The forced alignment step to generate phonemes only requires a decent pronunciation lexicon in that language or accent, which ideally covers a large part of the vocabulary, and an acoustic model that is pre-trained or trained from scratch (which does not require a large amount of speech data \cite{mcauliffe17_interspeech}). We postulate that the impact of the aligner is rather limited with respect to the data size for finetuning the LLM, but this could be further analyzed.
We have effectively applied our method to multiple datasets and to encoders of three types of speech models, i.e. a multilingual AED (attention-based encoder-decoder) model in Sec.~\ref{sec:100h} and Sec.~\ref{sec:ls960h}, an SSL model in Sec.~\ref{sec:ls960h} and a CTC-regularized monolingual AED model in Sec.~\ref{sec:qual}, showcasing it's broad applicability. While in some cases the dedicated ASR model baselines still attain lower WERs than the speech-LLM (as is a general trend in curent research),  we note that speech-LLMs have broader application and more benefits besides pure ASR tasks.

Phoneme-first decoding also provides explainability to the user, as the phonemes can indicate which sounds were recognized poorly by the LLM or the speech encoder, or similarly, which words were pronounced poorly by the user. As such, this work lays the ground for speech assessment tasks where a phonemic explanation is required. This two-step prediction comes at the expense of slower inference, since the output sequence is longer. It also requires (low-rank) adaptation of the LLM, which is often done anyway to better model spoken language and learn speech instruction tasks \cite{fathullah24_icassp, tang24_iclr, wu23_asru}, and could possibly leverage textual G2P data for improved pronunciation modeling. Finally, a more fine-grained perception of speech features and phonetic properties could be advantageous for acoustic matching of contextual bias words \cite{qiu23_asru, futami24_icassp}.

\section{Conclusion}
\noindent We have proposed a novel method to train speech-augmented LLMs by first predicting the phonemes with the LLM and then the words. We have shown that phoneme-first decoding can improve ASR, provides explainability, and yields phonemically-grounded transcripts.

\bibliographystyle{IEEEbib}
\bibliography{refs}

\end{document}